\documentclass[a4paper]{jpconf}
\usepackage{graphicx}
\usepackage[utf8]{inputenc}
\usepackage{amsmath,amssymb}
\usepackage{hyperref}
\usepackage[english]{babel}
\usepackage{color}
\usepackage[colorinlistoftodos]{todonotes}
\usepackage{subfigure}
\usepackage{wrapfig}
\usepackage{cite}

\begin{document}
\title{Prediction of the higher-order terms based on Borel resummation with conformal mapping}

\author{M.V.~Kompaniets}
\address{St. Petersburg State University,
7/9 Universitetskaya nab., 
St. Petersburg, 199034 
Russia.} 
\ead{m.kompaniets@spbu.ru}

\begin{abstract}

In this paper we discuss the method of the resummation of the asymptotic series suggested by Kazakov et al. in \cite{Kazakov:1978ey,Kazakov:1980rd} and predictions of the higher order terms based on this approach. An application of this method to the  $\varphi^4$ model is discussed.
\end{abstract}

\section{Introduction}
Motivated by the recent six loop computations of the field anomalous dimension in $\varphi^4$ model~\cite{6loopg2}  ($\varepsilon$-expansion)  and upcoming calculations of the six loop beta function in this model~\cite{bkpinprep}, we discuss in this paper the method of the resummation of the asymptotic series with factorially growing coefficients suggested  by Kazakov  et al. in~\cite{Kazakov:1978ey,Kazakov:1980rd}. This method is based on the Borel transform with a conformal mapping~\cite{Kazakov:1978ey,Kazakov:1980rd,cm1,zinn-cm-2,zinn-cm-3} and utilizes information about high order asymptotics (HOA)\cite{lipatov} as well as information about the  asymptotic behavior at large values of the coupling constant.

\section{General considerations}
Let us consider a quantity $A(g)=\sum_n A_n g^n$ which is defined as  an infinite series with factorially growing coefficients. The procedure of the Borel resummation has the following steps: first, one should find a Borel image $B(x)=\sum_n B_n x^n$ \eqref{inv_borel} (coefficients $B_n$ will have no factorial growth), then sum it up and, at last, perform the inverse Borel transform \eqref{inv_borel} to get the resummed value $A^B(g)$. Usually, a series defined by coefficients $B_n$ has finite radius of convergence, thus to define $B(x)$ on $x \in [0,\infty)$ one needs to perform an analytical continuation for $B(x)$ outside of the convergence radius.

The Borel transform can be defined as follows:
\begin{equation}
A^B(g) = \int_0^{\infty}{dt}\; e^{-t} t^{c_1-c_2}\left(t\frac{\partial}{\partial t}\right)^{c_2} \sum_n B^{(c_1,c_2)}_n (gt)^n\;, \quad B^{(c_1,c_2)}_n \equiv \frac{A_n}{\Gamma(n+1+c_1-c_2) n^{c_2}}\;.
\label{inv_borel}
\end{equation}
Usually the Borel transform with $c_2=0$ is used as the most simple one, but in some cases the transform with $c_2\neq 0$ gives a better convergence (see \cite{Kazakov:1978ey,Kazakov:1980rd})

 Working within the perturbative approach, the quantity $A(g)$ is known only up to some finite order of the perturbation theory:
\begin{equation}
A^{(N)}(g) = \sum\limits_{n=0}^N A_n g^n\;.
\label{AN}
\end{equation}
One can easily see that for a finite number of terms~\eqref{AN}, the Borel resummation described above simply reproduces the initial series $A^{(N)}(g)$. Thus to get some nontrivial result we need to make some assumptions about the higher order terms ($A_n$, $n>N$).
This is the most crucial step in the procedure of the Borel resummation, because a good choice (which will most accurately reproduce original (may be not known) series) will lead to a better convergence of the resummation procedure, while bad choice may lead to completely inconsistent results.

Knowing only first $N$ terms of the series \eqref{AN}, continuation of the series to large $N$ is a very ambiguous procedure, for example, one may use Pade approximants for that \cite{pade,bnm78}: 
\begin{equation}
A^{Pade-Borel} =  \int_0^{\infty}{dt}\; e^{-t} t^{c_1-c_2}\left(t\frac{\partial}{\partial t}\right)^{c_2} B(gt)\;, \quad
B(x)= \frac{P_L(x)}{P_M(x)}\;.
\end{equation}
Polynomials $P_L(x)$ and $P_M(x)$ are chosen in such a way that the initial terms of the series expansion of $B(x)$ reproduce the Borel image \eqref{inv_borel} of our series \eqref{AN}. For the series considered in $\varphi^4$ model and similar models, results obtained by Pade-Borel resummation are highly dependent on the choice of the approximant ($L$, $M$), and sometimes the approximant has a singularity in the integration domain, which prevents one from performing the inverse Borel transformation (see e.g. \cite{sokolov95,sokolov2000}).

To make results more reliable one needs to incorporate into the reconstructed series all the  information we know about it~\cite{nalimov2009borel}. For example, for the  $\varphi^4$ model (and some other models) we know high order asymptotics (HOA) for $A_n$ \cite{lipatov}:
\begin{equation}
A_n \sim (-1)^n \;n! \; n^{b_0} \;a^n (1+{\cal O}(1/n))\;.
\label{hoa}
\end{equation}
Parameters $a$ and $b_0$ are determined from the instanton analysis (see \cite{lipatov}).
To incorporate this information into the resummation procedure the method of the Borel transform with conformal mapping was developed \cite{cm1,Kazakov:1978ey,zinn-cm-2,Kazakov:1980rd,zinn-cm-3}.
In this method function $B(x)$ is defined as follows:
\begin{equation}
B(x) = \sum\limits_{n=0}^N C_n w(x)^n\;, \qquad \mbox{where}\;\; w(x) = \frac{\sqrt{1+ax}-1}{\sqrt{1+ax}+1}\;,
\label{B_CM}
\end{equation}here  $a$ is the parameter of HOA \eqref{hoa}, and $b_0$ fixes the value of $c_1=b_0+3/2$ in \eqref{inv_borel} to fix HOA of the reconstructed series  in accordance with \eqref{hoa}. And again, expansion  of the $B(x)$ must reproduce the initial coefficients of the  Borel image of the original series \eqref{AN}. The variable transformation in \eqref{B_CM} ensures that the reconstructed series has an analytical continuation in the whole integration domain.
This resummation method gives more reliable results for the resummed values (see e.g.\cite{Zinnbook,Pelissetto2002}), but still has some disadvantages: despite the fact that the series reconstructed in such a way reproduces the initial part of the original series \eqref{AN} and  corresponding  HOA \eqref{hoa}, the terms of the series with $n>N$ are reproduced incorrectly. Actually this means that we resum a series different than the original one (of course if  $N$ is big enough this procedure must converge to a correct value,  but given  maximum 6 terms  \cite{6loopg2} the correct choice of the function $B(x)$ is very important).

The next step was performed in the papers of Kazakov et al.~\cite{Kazakov:1978ey,Kazakov:1980rd}: the function $B(x)$ in \eqref{B_CM} was modified in such a way that it reproduces not only initial terms and HOA, but also the resummed function has the same asymptotic behavior at $g\to\infty$ as the initial function $A(g)$:
\begin{equation}
B(x) = \left(\frac{x}{w}\right)^\nu \left(\sum\limits_{n=0}^N  C_n w(x)^n\right), \qquad \mbox{where}\;\; w(x) = \frac{\sqrt{1+ax}-1}{\sqrt{1+ax}+1}\;,
\label{B_KST}
\end{equation}
obviously,  the parameter $\nu$  governs the behavior of the resummed function at $g\to\infty$ (in \eqref{inv_borel} $x=gt$)

In~\cite{Kazakov:1978ey,Kazakov:1980rd} it was shown that for the series with known asymptotic behavior at $g\to\infty$ the most correct resummed values are obtained when the parameter $\nu$  is chosen close to the asymptotic. Moreover, in this case the expansion of the $B(x)$ reproduces the terms with $n>N$ with high accuracy. Thus we can state that with this approach we resum the series which is really close to the original one.

Let us illustrate this on the expansion of the following integral:
\begin{equation}
A(g) = \int\limits_0^\infty dx\; e^{-x^2-g x^4}=\sum\limits_{n=0}^\infty A_n g^n\;,\quad \left(A_0=\frac{\sqrt{\pi}}{2}, \;\; A_n = \frac{(-1)^n \sqrt{\pi}(4n-1)!}{(4)^{2n} n! (2n-1)!}, \; (n>0)\right)\;.
\label{Ag}
\end{equation}
High order asymptotic of this series is given by \eqref{hoa} with $a=4$ and $b_0=-1$ and $A(g) \sim g^{-1/4}$, when $g\to\infty$.

\begin{figure*}[t]
    \centering
    \subfigure[Relative prediction error $\xi_N={(A^{(N)}_6-A_6)}/{A_6}$ as a function of $\nu$ for different values of $N$.]{\label{gx4_pred}\includegraphics[width=6.8cm]{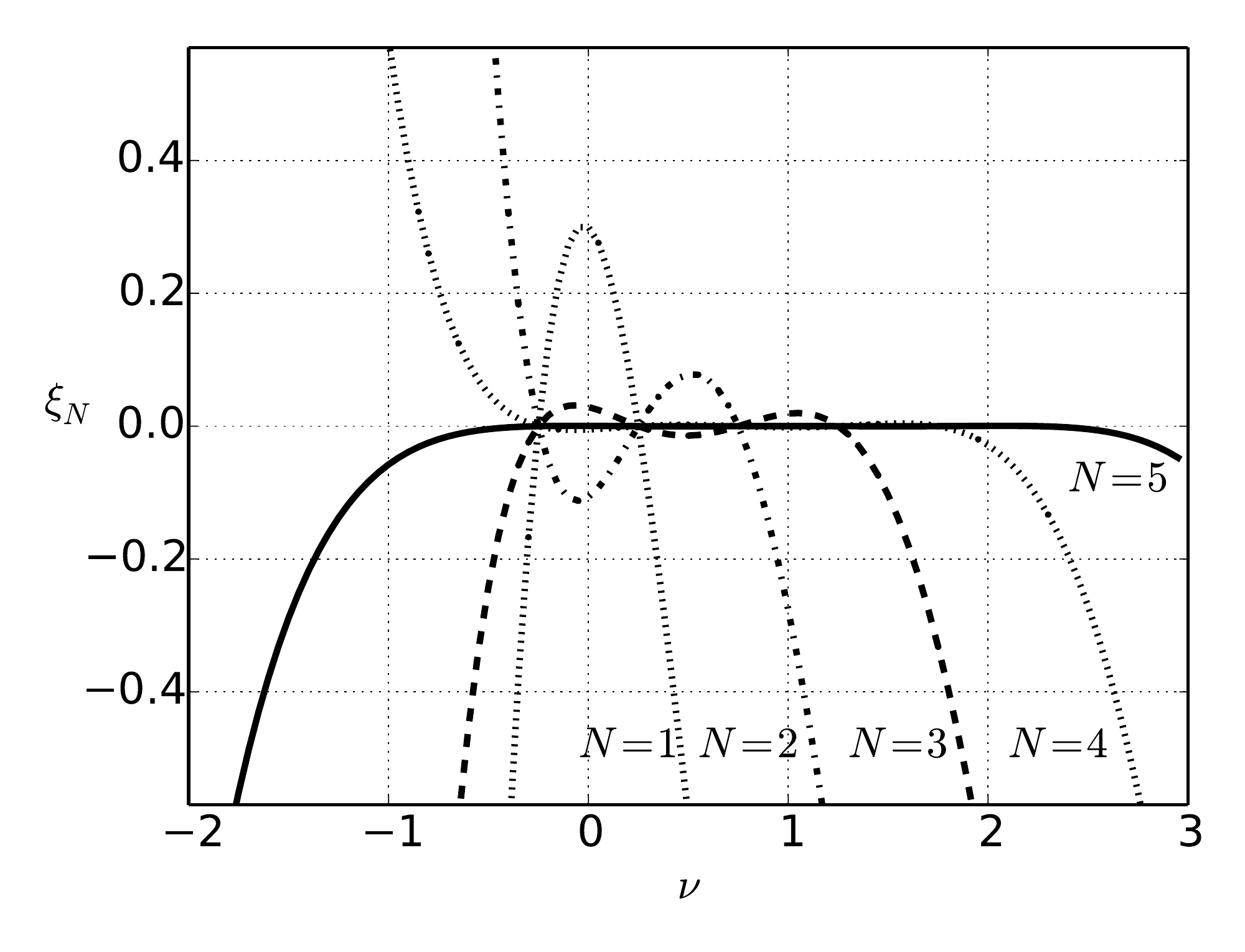}}~~~~~
    \subfigure[The value of the resummed series $A^{(N)}$ at $g=1$, for different $N$ and $\nu$.]{\label{gx4_res}\includegraphics[width=6.8cm]{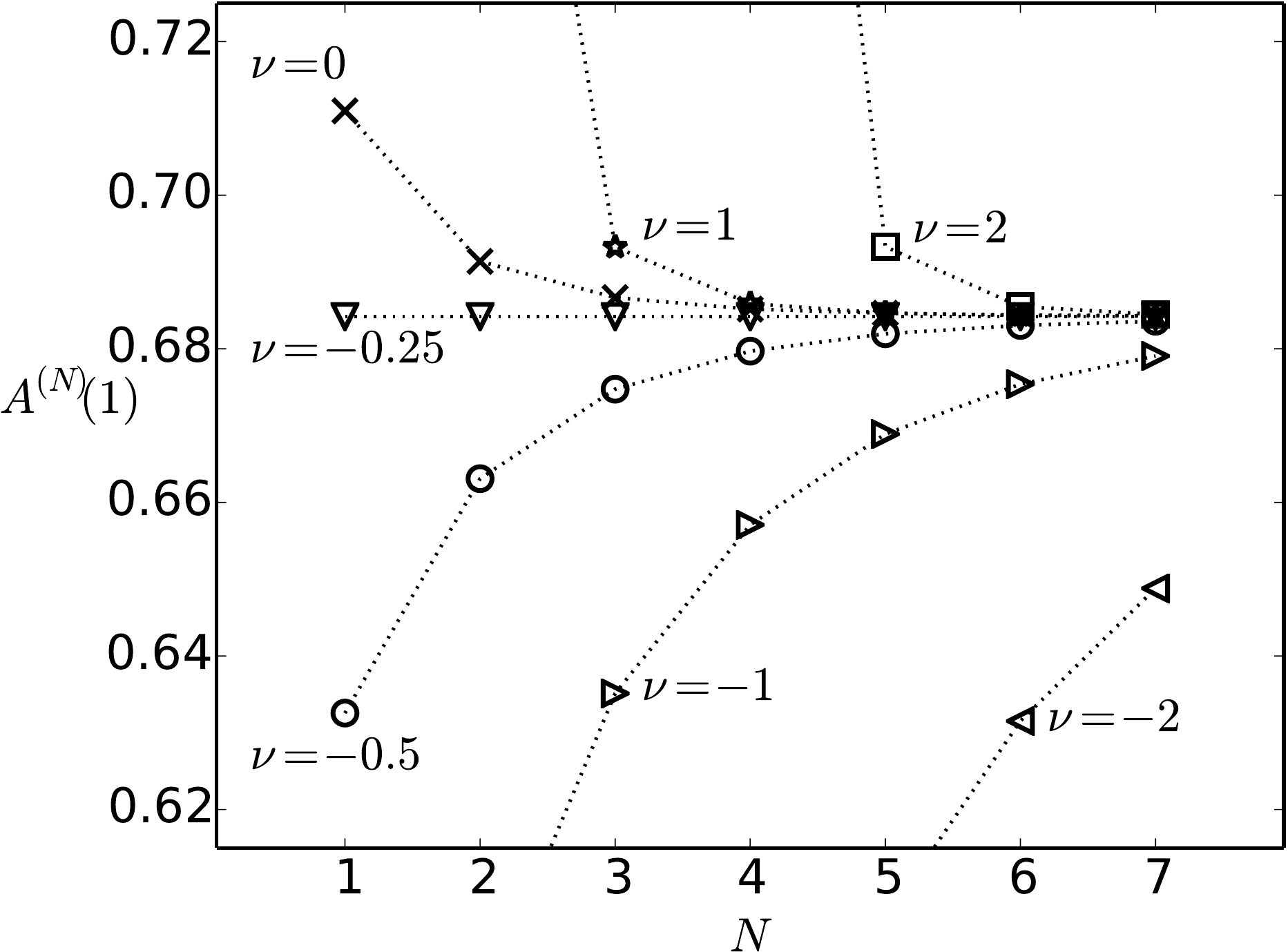}}	
    \caption{Zero dimensional $\varphi^4$ model, see Eq.~ \eqref{Ag}.}
    \label{gx4_ab}
\end{figure*}

If we calculate the function $B^{(N)}(x)$ from \eqref{B_KST} for the first $N$ terms of the expansion of the integral \eqref{Ag}, we can expand it back in $x$ and reconstruct the series, which we are actually going to resum $B^{(N)}(x)=\sum A_n^{(N)}x^n$. From Fig.\ref{gx4_ab} one can see that a better convergence of the resummation procedure (Fig.\ref{gx4_res}) appears for such values of $\nu$ where predictions of higher-order terms are most accurate (Fig.\ref{gx4_pred}).  From Fig.\ref{gx4_pred} one can see that large negative and positive values of $\nu$ give us completely incorrect predictions, while starting from $\nu=-1/4$ we have an area with good predictions. One can easily  see that the representation \eqref{B_KST} with $\nu=-1/4$  exactly reproduces the coefficients of the initial series~\eqref{Ag} thus giving a resummed result starting from $N=0$. Of course, this fact can be explained by the simplicity of the example considered, and in more complicated cases (like $\varphi^4$ model) it is not the case.
Any way, using this example we can investigate the stability of this resummation procedure and it's convergence. 

\section{Resummation of the $\varphi^4$ model in $D=4-2\varepsilon$ space dimensions}
\begin{figure*}[t]
    \centering
    \subfigure[The relative prediction error $\xi_N^\beta$ as a function of $\nu$, for the different values of $N$. ]{\label{phi4_pred}\includegraphics[width=6.8cm]{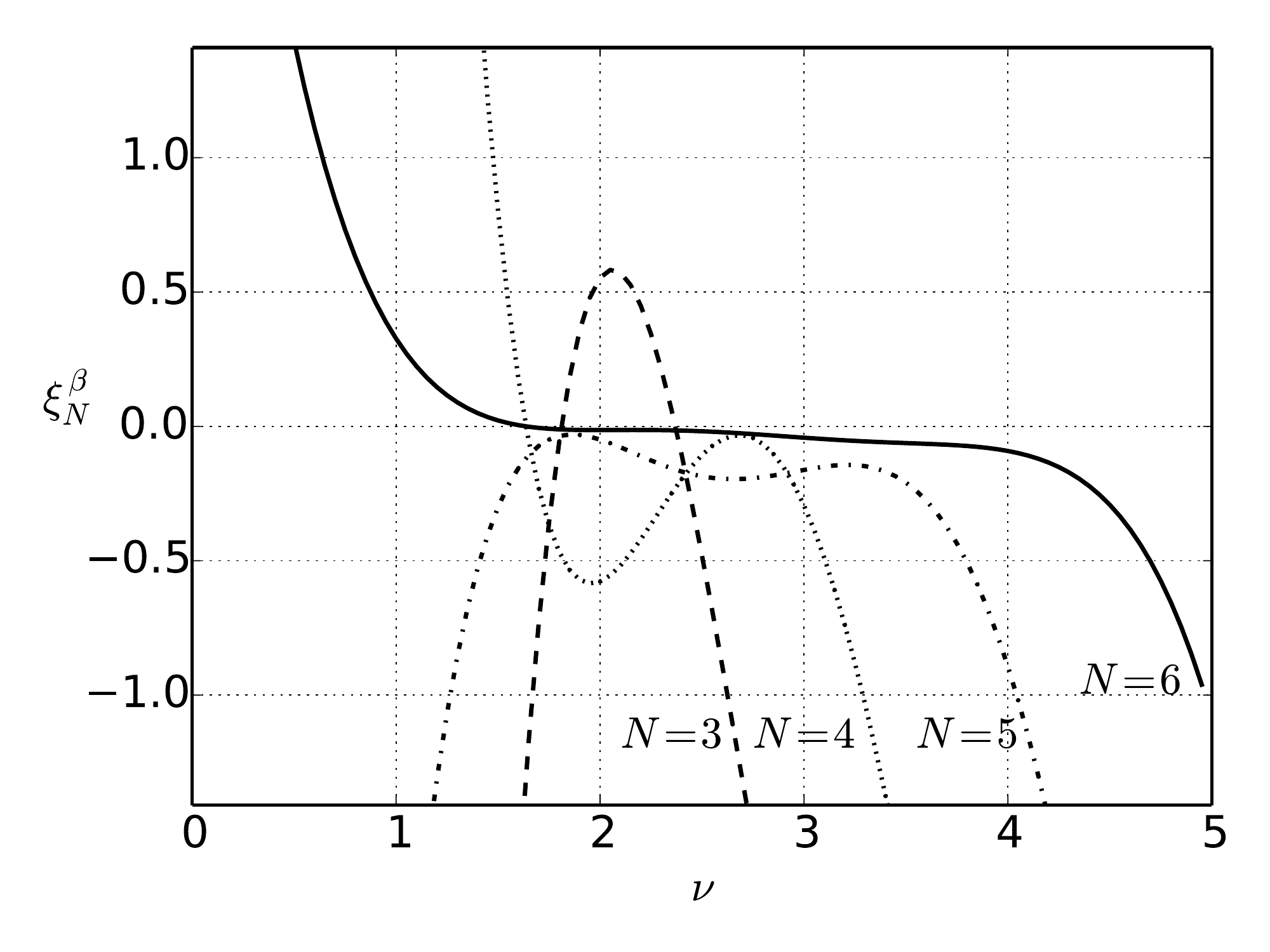}}~~	
    \subfigure[The value of the resummed beta function $\beta^{(N)}$ at $g=1$, for different $N$ and $\nu$.]{\label{phi4_beta}\includegraphics[width=6.8cm]{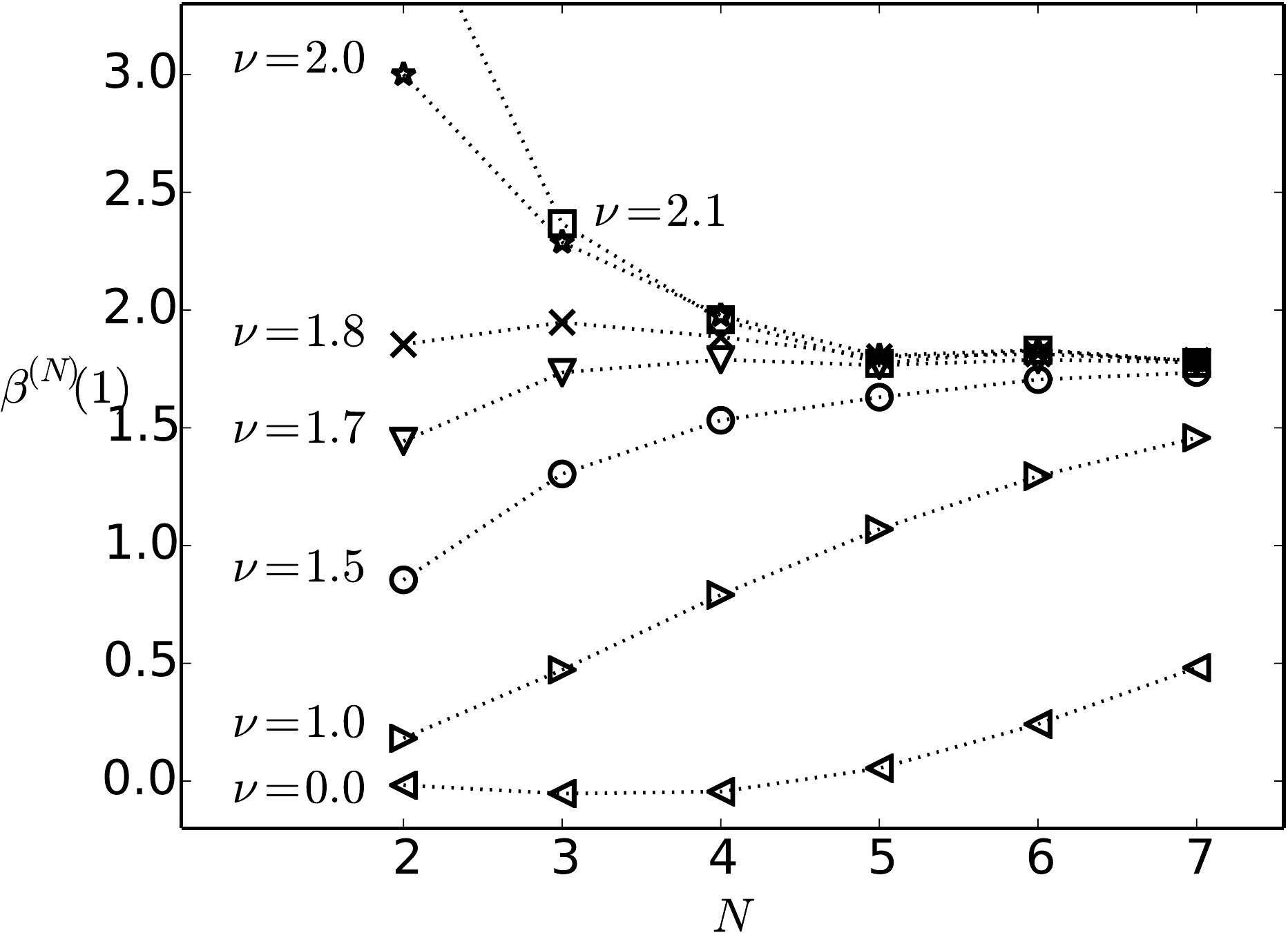}}	
    \caption{$\varphi^4$ model ($D=4-2\varepsilon$, MS-scheme).}
    \label{phi4_ab}
\end{figure*}

The situation similar to the described  above appears for the $\varphi^4$ model beta-function (see Fig.\ref{phi4_ab}). In this model the exact result is not known as well as coefficients of the beta-function for arbitrary number of loops. But we can verify the predictive power of this approach by checking predictions for the highest known term from lower ones as well as investigating  the stability of the resummation procedure like in zero dimensional model.

For this model the plot with the relative prediction error $\xi^\beta_N={(\beta^{(N)}_M-\beta_M)}/{\beta_M}$ (Fig.\ref{phi4_pred}) has a similar form, of course there is no exact intersection as in Fig.\ref{gx4_pred}.
One can see that the standard Borel transform with conformal mapping \eqref{B_CM} (which corresponds to $\nu=0$) gives completely incorrect predictions, while near $\nu=2$ predictions are almost correct. In papers \cite{Kazakov:1978ey,Kazakov:1980rd} the range $1.7<\nu<2.2$ was obtained for the beta-function and the value $\nu=2$ was recommended to use for the resummation.

The same situation appears for the convergence of the resummation procedure: on  Fig.\ref{phi4_beta} the convergence of the resummation of the beta function $\beta^{(N)}(g)$ at $g=1$ is shown for the different values of the parameter $\nu$ and the number of terms $N$ taken into account (note that e.g. $N=3$ corresponds to the  two loop approximation; the preliminary value of the six loop term for the beta function is taken from \cite{bkpinprep}).
\begin{wrapfigure}{r}{0.43\textwidth}
\includegraphics[width=6.8cm]{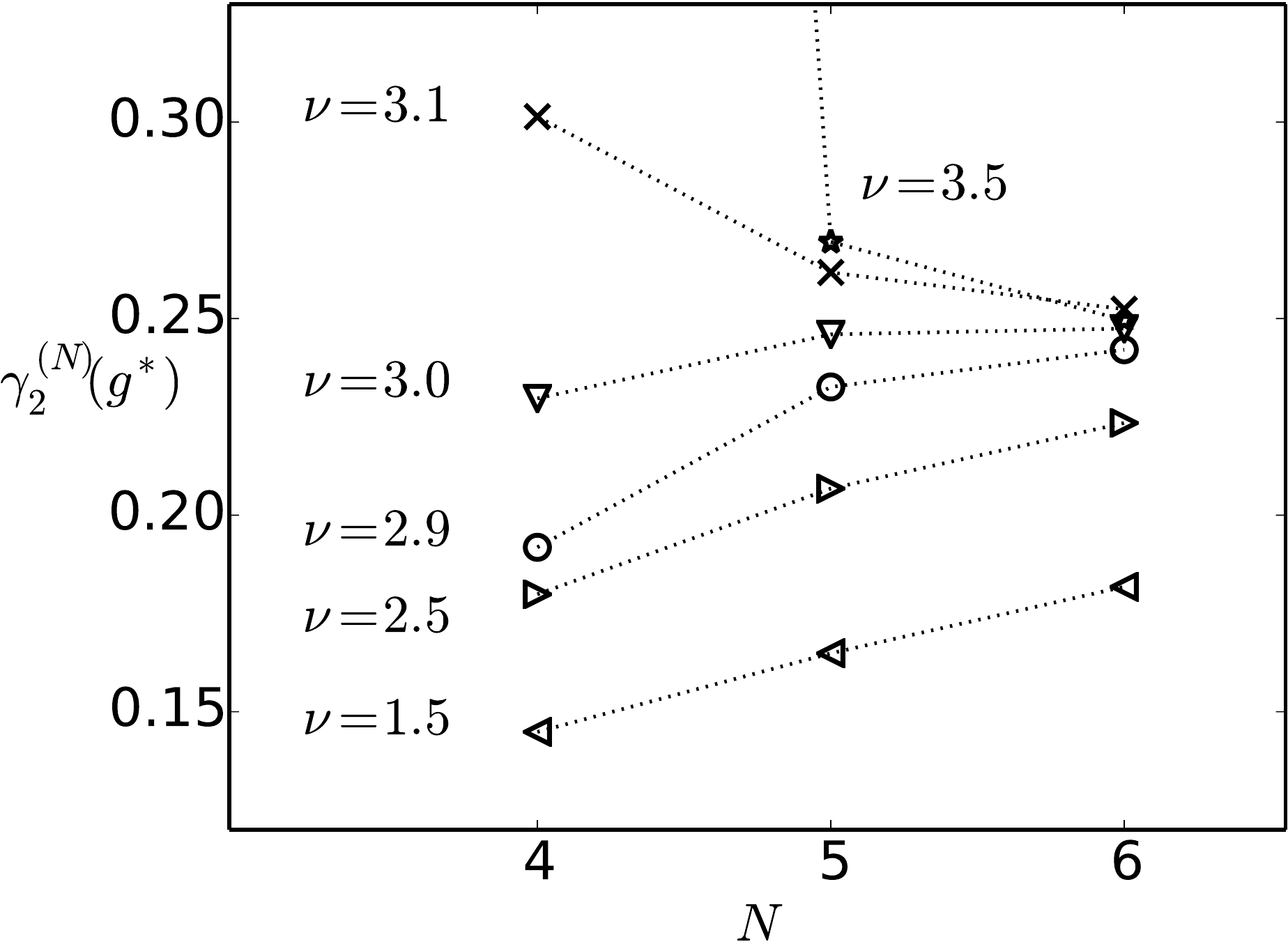}
\caption{Resummed values of the Fisher exponent $\eta\equiv 2\gamma_\varphi(g^*)$ at $g^*=1.155$ (fixed point at $D=2$) for the different values of $N$ and $\nu$.}
\label{phi4_eta}
\end{wrapfigure}
The $\nu$ dependence of the resummed beta function  $\beta^{(N)}(g=1)$\footnote{The similar behavior appears for any reasonable(not very large) $g$. } has a form similar to the zero dimensional model (Fig.~\ref{gx4_res}). And again we can see that the value of  $\nu$ with the most correct predictions ($\nu \simeq 1.8$) gives us the most stable resummed value of the beta-function, therefore by analogy with the zero dimensional model we should use this value in the resummation procedure. Also it should be noted that the resummation procedure for  the field anomalous dimension for $N=2$ and $3$ leads to completely inconsistent results for any $\nu$,  but starting from the $N=4$ shows the same properties (see Fig.~\ref{phi4_eta}), and according to this plot for the resummation of the field anomalous dimension the value of the parameter $\nu$ must be chosen as $\simeq 3$.

Performing the resummation analogous to the one made in \cite{6loopg2} but with $\nu_\beta=1.8$\footnote{In \cite{6loopg2} the value $\nu_\beta=2$ was used.} and $\nu_\eta=3$ we arrive to the  resummed exponents shown in Table~\ref{table:resummed2}.

\begin{table}[h!]
\caption{Resummation result ($\nu_\beta=1.8$) for the Fisher exponent $\eta$ for different number of loops taken into account. Expected values are $0.25$  and $0.0364(5)$~\cite{Pelissetto2002}, for $D=2$ and $D=3$.}
\label{table:resummed2}
\centering
\begin{tabular}{c c c c c c c c c} 
\br
Loops $\beta/\gamma_\varphi$& 3/4& 3/5 & 3/6& 4/4 & 4/5 & 4/6 & 5/5 & 5/6 \\[0.5ex] 
 \mr
$D=2$ &0.1952 & 0.2079& 0.2090& 0.2233& 0.2388& 0.2403& 0.2341& 0.2354 \\[0.5ex] 
$D=3$ &0.03351& 0.03406& 0.03410& 0.03556& 0.03623& 0.03628& 0.03599& 0.03604\\
\br
 \end{tabular}

\end{table}

Comparing with the table in \cite{6loopg2} one can see that the value $\nu_\beta=1.8$ gives us better convergence of the resummation procedure as well as better predictions for the beta function coefficients.
For a five loop term of the beta-function predicted by a 4 loop beta function we have ($M=6$, $N=5$) $\beta_6^{(5)} = 2821.65$, while the exact value is $\beta_6 = 2848.57...$ which is only $1\%$ greater than the prediction, in \cite{Kazakov:1978ey,Kazakov:1980rd,phi4g4} value $\simeq 2808$ was obtained for $\beta_6^{(5)}$. 
For a six loop term on top of a five loop beta function we have ($M=7$, $N=6$) $\beta_7^{(6)} = -34393.3$. Of course, we didn't expect that a six loop term of the beta function will exactly fit this prediction, moreover it is expected that the absolute value of the six loop term is a bit greater than that of prediction.

\section{Conclusions}
Summarizing, we have shown that it is possible to choose such a form of the analytical continuation of the Borel image which will most accurately reproduce higher order terms of the series. This will provide a better convergence of the resummation procedure as well as predictions of the higher order terms. Despite the fact that some arguments for the  choice of the $B(x)$ in Eq.~\eqref{B_KST} and the particular value of $\nu$ for $\varphi^4$ model are presented, this question is still open. 
It  is still necessary to find more strict arguments for the choice of $B(x)$ and $\nu$.

\section*{Acknowledgments}
I am grateful to  { L~Ts~Adzhemyan},{ D~I~Kazakov}, {M~Yu~Nalimov}, A~I~Sokolov and {E~Zerner-K{\"a}ning} for fruitful discussions, to ACAT’16 Organizing Committee, and in particular to I~Kondrashuk (Univ. of Bio-Bio, Chill{\'a}n), L~Salinas (UTFSM, Valparaiso) and Y~Schroder (Univ. of Bio-Bio, Chill{\'a}n) for support and hospitality, to A~L~Kataev for the support in the process of the preparation of this talk. The work was supported  by Saint-Petersburg State University (project 11.38.185.2014).  

\section*{References}
\medskip
\bibliography{paper}

\providecommand{\newblock}{}
\begin{thebibliography}{10}
\expandafter\ifx\csname url\endcsname\relax
  \def\url#1{{\tt #1}}\fi
\expandafter\ifx\csname urlprefix\endcsname\relax\def\urlprefix{URL }\fi
\providecommand{\eprint}[2][]{\url{#2}}

\bibitem{Kazakov:1978ey}
Kazakov D, Shirkov D and Tarasov O 1979 {\em Theor. Math. Phys.\/} {\bf 38} 9

\bibitem{Kazakov:1980rd}
Kazakov D and Shirkov D 1980 {\em Fortsch. Phys.\/} {\bf 28} 465

\bibitem{6loopg2}
Batkovich D, Chetyrkin K and Kompaniets M 2016 {\em Nuclear Physics B\/} {\bf
  906} 147 -- 167 (\textit{Preprint} \eprint{hep-th/1601.01960})

\bibitem{bkpinprep}
Kompaniets M and Panzer E \textit{in preparation}

\bibitem{cm1}
Loeffel J 1996 {\em Workshop on Pade approximants\/} ed Bessis D, Gilewicz J
  and Merry P (ACM Press)

\bibitem{zinn-cm-2}
Le~Guillou J~C and Zinn-Justin J 1977 {\em Phys. Rev. Lett.\/} {\bf 39}(2)
  95--98

\bibitem{zinn-cm-3}
Le~Guillou J~C and Zinn-Justin J 1980 {\em Phys. Rev. B\/} {\bf 21}(9)
  3976--3998

\bibitem{lipatov}
Lipatov L 1977 {\em J. Exptl. Theoret. Phys.\/} {\bf 72} 411

\bibitem{pade}
Baker G 1975 {\em Essentials of Pade Approximants\/} (Academic, New York)

\bibitem{bnm78}
Baker G~A, Nickel B~G and Meiron D~I 1978 {\em Phys. Rev. B\/} {\bf 17}(3)
  1365--1374

\bibitem{sokolov95}
Antonenko S~A and Sokolov A~I 1995 {\em Phys. Rev. E\/} {\bf 51}(3) 1894--1898

\bibitem{sokolov2000}
Orlov E and Sokolov A 2000 {\em Physics of the Solid State\/} {\bf 42}
  2151--2158

\bibitem{nalimov2009borel}
Nalimov M~Y, Sergeev V~A and Sladkoff L 2009 {\em Theor. Math. Phys.\/} {\bf
  159} 499--508

\bibitem{Zinnbook}
Zinn-Justin J 2002 {\em Quantum Field Theory and Critical Phenomena\/} (Oxford:
  Clarendon Press)

\bibitem{Pelissetto2002}
Pelissetto A and Vicari E 2002 {\em Physics Reports\/} {\bf 368} 549--727
  (\textit{Preprint} \eprint{cond-mat/0012164})

\bibitem{phi4g4}
Chetyrkin K~G, Gorishny S~G, Larin S~A and Tkachev F~V 1983 {\em Phys. Lett.
  B\/} {\bf 132} 351--354

\end{thebibliography}

\end{document}